# Topological Phase Transition in Quasi-One-Dimensional Bismuth Iodide Bi$_4$I$_4$

W. X. Zhao[1], M. Yang[2,3], X. Du[1], Y. D. Li[1], K. Y. Zhai[1], Y. Q. Hu[1], J. F. Han[4,5], Y. Huang[4,5], Z. K. Liu[6,7], Y. G. Yao[4,5], J. C. Zhuang[2,3], Y. Du[2,3*], J. J. Zhou[4*], Y. L. Chen[6,7,8*], and L. X. Yang[1,9,10*]

[1]State Key Laboratory of Low Dimensional Quantum Physics, Department of Physics, Tsinghua University, Beijing 100084, China.

[2]School of Physics, Beihang University, Beijing 100191, China.

[3]Centre of Quantum and Matter Sciences, International Research Institute for Multidisciplinary Science, Beihang University, Beijing 100191, China.

[4]Centre for Quantum Physics, Key Laboratory of Advanced Optoelectronic Quantum Architecture and Measurement (MOE), School of Physics, Beijing Institute of Technology, Beijing 100081, China.

[5]International Center for Quantum Materials, Beijing Institute of Technology, Zhuhai 519000, China

[6]School of Physical Science and Technology, ShanghaiTech University and CAS-Shanghai Science Research Center, Shanghai 201210, China.

[7]ShanghaiTech Laboratory for Topological Physics, Shanghai 200031, China.

[8]Department of Physics, Clarendon Laboratory, University of Oxford, Parks Road, Oxford OX1 3PU, UK.

[9]Frontier Science Center for Quantum Information, Beijing 100084, China.

[10]Collaborative Innovation Center of Quantum Matter, Beijing China.

*e-mail: LXY: lxyang@tsinghua.edu.cn. YD: yi_du@buaa.edu.cn; JJZ: jjzhou@bit.edu.cn; YLC: yulin.chen@physics.ox.ac.uk;

**Abstract**

**The exploration of topological quantum materials and topological phase transitions is at the forefront of modern condensed matter physics. Quasi-one-dimensional (quasi-1D) bismuth iodide $Bi_4I_4$ exhibits versatile topological phases of matter including weak topological insulator (WTI) and higher-order topological insulator (HOTI) phases with high tunability in response to external parameters. In this work, performing laser-based angle-resolved photoemission spectroscopy with submicron spatial resolution (micro-ARPES), we comprehensively investigate the fine electronic structure and topological phase transition of $Bi_4I_4$. Our examination of the low-temperature α-phase reveals the presence of an energy gap on the (100) surface, providing spectroscopic evidence for the HOTI phase. Conversely, the high-temperature β-$Bi_4I_4$ harbors a gapless Dirac fermion on the (100) surface alongside gapped states on the (001) surface, thereby establishing a WTI phase. By tracking the temperature evolution of the (100) surface states, we unveil a thermal hysteresis of the surface gap in line with the α-β structural phase transition. Our findings elucidate the topological properties of $Bi_4I_4$ and directly evidence a temperature-induced topological phase transition from WTI to HOTI, which paves the way to potential applications based on the room-temperature topological phase transition in the quasi-1D topological quantum material.**

## INTRODUCTION

Topological quantum phase transitions (TQPTs) represent a fascinating type of phase transitions characterized by global topological invariants, which diverge from the conventional symmetry-breaking phase transitions described by local order parameters. TQPTs not only hold significant scientific implications for elucidating topological properties under external perturbations[1-5] but also pave an avenue for the utilization of topological quantum materials with high tunability in future electronic and spintronic devices[6-9]. Throughout a TQPT, the ground-state wavefunction evolves with a controlling parameter such as strain, chemical substitution, electric field, and temperature[10-17], typically associated with a change in the energy gap of the boundary states.

Recently, quasi-one-dimensional (quasi-1D) bismuth halides $Bi_4X_4$ ($X$ = Br, I) have garnered great research attention owing to the abundant topological quantum phases in this family compounds, including strong/weak topological insulator[18-20], higher-order topological insulator[21-23], and Weyl semimetal[24]. These intriguing topological quantum phases are sensitive to the stacking sequence and interlayer coupling of $Bi_4X_4$ layers constituting the bulk crystals[25-27]. By manipulating external conditions, e.g., tuning temperature, applying strain or chemical substitution, TQPT can be established and feasibly controlled.

Taking $Bi_4I_4$ as an example, monolayer $Bi_4I_4$ consists of arrays of chains along the $b$ axis, which has been verified as near a 2D quantum spin Hall (QSH) insulator/normal insulator phase boundary[28-30]. Bulk α/β-$Bi_4I_4$ share the same chain structure except that the α-phase comprises two $Bi_4I_4$ chains relatively shifting along the chain direction in a unit cell (Figs. 1a, b). Consequently, the interaction between the QSH edge states in each $Bi_4I_4$ monolayer is distinctive in the two phases. For β-$Bi_4I_4$, the monolayers stack along $c$ with the metallicity of the side surface preserved by the symmetry, realizing a WTI phase[4,12,31-34]. By contrast, the adjacent QSH edge states are in different chemical environments in α-$Bi_4I_4$. Therefore, the hybridization between adjacent QSH edge states induces gapped (100) surface states (Figs. 1b, c). Remarkably, for particular stacking sequences, gapless modes, namely topological hinge states, survive at the hinges of the crystal as enforced by the higher-order bulk-boundary correspondence (see the red dots in

Fig. 1b)[27,35]. Since the dimensional difference between the bulk and conducting boundary mode is larger than 1, this intriguing topological state is classified as a HOTI. Upon increasing temperature, α-Bi$_4$I$_4$ undergoes a structural transition to β-Bi$_4$I$_4$ near $T_S$ ~ 300 K, establishing a temperature-induced TQPT[36].

Angle-resolved photoemission spectroscopy (ARPES), serving as a powerful tool to elucidate the electronic structure of single crystals, has been performed to unravel the band structure of bismuth halides[26,37-40]. While high-resolution laser-ARPES experiments have shown evidence for the HOTI phase in α-Bi$_4$Br$_4$, the investigation of Bi$_4$I$_4$ is more intricate[35,41,42]. Gapless (100) surface states have been observed in β-Bi$_4$I$_4$[35,42], confirming its non-trivial topological property. However, there are still debates on the existence of gapless surface states on the (001) surface, making it elusive whether β-Bi$_4$I$_4$ is a WTI or strong topological insulator (STI). As for α-Bi$_4$I$_4$, although a gapped (100) surface state has been detected at the $\bar{\Gamma}_{100}$ point[35] in accordance with the theoretical prediction, further confirmation is still required to ascertain whether the side surface states are fully gapped over the whole surface Brillouin zone (especially at the $\bar{Z}$ point) and conclusively identify the HOTI phase. Moreover, only subtle change of the width of ARPES spectra was observed across the structural transition[35], with the temperature-evolution of the fine electronic structure, especially the surface gap, yet to be investigated.

In this work, by performing laser-based micro-ARPES measurements with high energy and momentum resolution[37,43], we systematically investigate the temperature evolution of the electronic structure of Bi$_4$I$_4$. At low temperatures, we observe gaps of about 40 meV and 5 meV at the $\bar{\Gamma}_{100}$ and $\bar{Z}$ points on the (100) surface, confirming the HOTI phase of α-Bi$_4$I$_4$. With increasing temperature, we directly observe the evolution of the gapped surface states into a gapless Dirac fermion across $T_S$. The magnitude of the gap shows a thermal hysteretic behavior, confirming the first-order nature of the TQPT. Moreover, our measurements unveil that a fully gapped (001) surface of β-Bi$_4$I$_4$, which, together with the gapless topological surface states (TSSs) on the (100) surface, compellingly evidences the WTI phase of β-Bi$_4$I$_4$. Our experimental observations are in excellent agreement with *ab-initio* calculations and provide more systematic insights into the TQPT of quasi-1D Bi$_4$I$_4$.

**RESULTS AND DISCUSSION**

Figure 2a shows the ARPES spectrum of α-Bi$_4$I$_4$ at the $\bar{\Gamma}_{100}$ point on the (100) surface. We resolve a W-shape conduction band bottom and an M-shape valence band top, characterizing the splitting feature and hybridization of the QSH edge states from adjacent Bi$_4$I$_4$ layers (Fig. 1c). According to the curvature spectra in Fig. 2c, the energy gap between the valence and conduction band is estimated to be about 31 meV (the peak-to-peak gap between the band crossing points at $\bar{\Gamma}_{100}$ is about 40 meV). At the $\bar{Z}$ point, the surface states exhibit a Dirac-cone-like dispersion with a gap of about 7 meV (Figs. 2b, d). Our measurements agree with the calculation in Figs. 2e and 2f. The gapped dispersion at $\bar{Z}$ can be further evidenced by fitting the Dirac dispersion as shown in Fig. 2b. The band dispersion is first extracted by fitting the momentum distribution curves (MDCs) to the Lorentzians (represented by the green dots). Then the extracted dispersion is fitted to the Dirac-fermion dispersion (yellow dashed lines), which gives an energy gap of about 5 meV at the Dirac point, in line with the curvature analysis in Fig. 2d.

Figures 2g and 2h show the laser-ARPES spectra of α-Bi$_4$I$_4$ measured on the (001) surface with their curvature plots shown in Figs. 2i and 2j, respectively. Around the $\bar{\Gamma}_{001}$ point, we observe a band gap over 395 meV without the conduction band observed. At the $\bar{M}$ point, we observe a direct band gap of about 170 meV. Both the experiments at $\bar{\Gamma}_{001}$ and $\bar{M}$ are well reproduced by the theoretical calculation of the bulk states, suggesting the absence of topological surface states on the (001) surface of α-Bi$_4$I$_4$. Therefore, our results on the (001) and (100) surfaces confirm the HOTI phase of α-Bi$_4$I$_4$, resembling the results of α-Bi$_4$Br$_4$[37]. Next, we examine the electronic structure of the (001) surface of β-Bi$_4$I$_4$ at 316 K. As shown in Figs. 3a and 3b, β-Bi$_4$I$_4$ exhibits a band gap of 330 meV (between the valence band top and $E_F$) and 125 meV (between the valence band top and the conduction band bottom) at the $\bar{\Gamma}_{001}$ and $\bar{M}$ points, respectively, similar to α-Bi$_4$I$_4$. The band dispersions and the band gaps are better visualized in the curvature plots of the data in Figs. 3c and 3d. Compared to α-Bi$_4$I$_4$, the valence and conduction band of β-Bi$_4$I$_4$ shift up slightly with a weak residual spectral weight of the conduction band at $E_F$. We do not observe any signature of topological surface states, which is in line with the WTI phase of β-Bi$_4$I$_4$ (see below).

In order to study the topological phase transition of bismuth iodide, we perform systematic ARPES measurements at different temperatures across $T_S$. At 80 K, the data at the $\bar{\Gamma}_{100}$ point (Fig. 4a) clearly shows a gap, consistent with the (100) surface states of α-Bi$_4$I$_4$ in Fig. 2a. With increasing temperature up to around 310 K, the energy interval between the surface valence and conduction band at the $\bar{\Gamma}_{100}$ point slightly decreases. The spectra also get blurred at high temperatures, which may indicate a preclude of crystal transformation related to the structural transition instead of thermal broadening effect[35]. Prominently, above $T_S$, the surface gap is suddenly suppressed and the dispersion shows a single gapless Dirac dispersion (Fig. 4g), which confirms the WTI phase of β-Bi$_4$I$_4$ together with the gapped spectra on the (001) surface (Fig. 3). Simultaneously, the spectrum becomes much sharper with the full width at the half maximum changed from about 0.01 Å$^{-1}$ (302 K) to 0.007 Å$^{-1}$ (316 K). This band sharpening of surface states is because of the bilayer splitting of α-Bi$_4$I$_4$ (Fig. 1c) and has been revealed by the previous experiments[35]. As the temperature decreases, the spectrum recovers at low temperatures and the surface gap reemerges (Figs. 4h-k).

The temperature evolution of the (100) surface gap extracted by double-Lorentzian fitting the EDC at $\bar{\Gamma}_{100}$ is summarized in Fig. 4l. It is worth noting that the surface gap exhibits a thermal hysteresis loop near the phase transition. Specifically, an α-phase characterization is maintained up to about 310 K in the heating process, which can be confirmed by features including surface gap, band sharpness, and band crossing energy mentioned above Figs. 4a-f. While in the cooling process, a β-phase spectrum with a sharp single Dirac cone is evident even at 295 K in Fig. 4i, in significant contrast with Fig. 4e detected at the same temperature, indicating that Bi$_4$I$_4$ undergoes a first-order structural phase transition with its topological phase varying from HOTI to WTI, in accordance with our *ab-initio* calculation.

For a long time, 2D QSH insulators, hosting symmetry-protected 1D edge states immune to backscattering, are expected to promote the development of dissipationless electronics and spintronics devices[4,5,44].

However, the complicated fabrication of 2D-material devices limits the exploration and application of the 1D edge states. By contrast, 1D topological quantum materials such as $Bi_4X_4$ can be relatively easily synthesized and naturally harbor highly directional topological boundary states with spin-momentum locking property, which are beneficial for the preservation of coherent quantum states. The quasi-1D nature of $Bi_4X_4$ further allows for the integration into nanowire configurations that is natural and ideal for device engineering. In particular, the novel HOTI phase with hinge states is promising for dissipationless transport, efficient charge-to-spin conversion, topological quantum computation based on the Majorana zero modes, and the realization of spin-triplet superconductivity[45-48]. Furthermore, the TQPT between HOTI and WTI phases in $Bi_4I_4$ bridged by a first-order structural transition provides a unique opportunity for the controllable manipulation of the topological boundary states near room temperature.

To conclude, we systematically investigate the electronic structure and its temperature evolution of $Bi_4I_4$. We observe a fully gapped (100) surface state of α-$Bi_4I_4$, supporting the identification of its HOTI phase. Our measurements at high temperatures confirm the WTI phase of β-$Bi_4I_4$ by detecting its gapped (001) band structure and gapless (100) TSS. More importantly, we detect the thermal hysteresis loop of the gap in the (100) surface states. Our study unravels the fine band structure of $Bi_4I_4$ with unprecedented precision and provides solid evidence on the TQPT between HOTI and WTI phases bridged by the structural transition. Our results provide a deeper scientific insight and pave the way to promising applications on the novel topological quantum properties of quasi-1D $Bi_4I_4$.

**METHODS**

High-quality $Bi_4I_4$ single crystals were synthesized using a solid-state reaction and chemical vapor transport method[49]. Figure 1g presents the cross-section image by atomic-resolved scanning transmission electron microscope, which confirms the crystal structure and the high quality of our samples. Laser-based μ-ARPES measurements with a sub-micron spatial resolution were conducted at Tsinghua University with the 7-eV laser generated by a $KBe_2BO_3F_2$ (KBBF) crystal and focused by an optics lens[43]. The samples were cleaved *in-situ* under UHV better than $1 \times 10^{-10}$ mbar and the data were collected by Scienta DA30L electron analyzer. The total energy and angular resolutions were set to 3 meV and 0.2°, respectively. First-principles calculations were carried out using the Vienna *ab-initio* simulation package (VASP)[50-53].

**DATA AVAILABILITY**

The data sets that support the findings of this study are available from the corresponding author upon request.

**ACKNOWLEDGEMENTS**

This work is funded by the National Key R&D Program of China (Grants No. 2022YFA1403100 and No. 2022YFA1403200) and the National Natural Science Foundation of China (Grants No. 12275148 and No. 62275061).

**AUTHOR CONTRIBUTIONS**

L.X.Y. conceived the experiments. W.X.Z. carried out ARPES measurements with the assistance of X.D., Y.D.L., K.Y.Z., Y.Q.H and Z.K.L. *Ab-initio* calculations were performed by J.J.Z. Single crystals were synthesized and characterized by M.Y., J.C.Z., and Y.D. The paper was written by W.X.Z, L.X.Y., and Y.L.C. All authors contributed to the scientific planning and discussion.

**COMPETING INTERESTS**

The authors declare no competing interests.

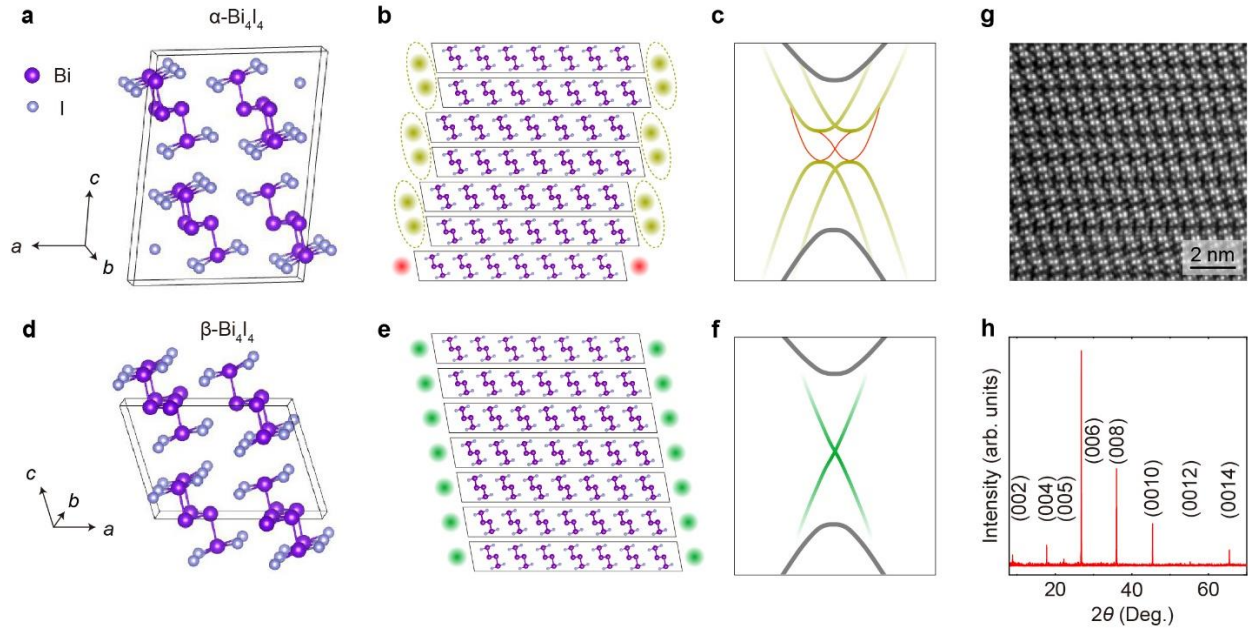

**Fig. 1 Topological properties and crystal structures of α/β-Bi4I4. a** The crystal structure of α-Bi4I4. The black lines indicate the unit cell. **b** Schematic illustration of the boundary states on the (100) surfaces. Each dot represents the QSH edge state of a Bi4I4 monolayer. The yellow-dot pair in an ellipse and red dots represent the quantum-hybridized (100) surface states and the topological hinge states, respectively. **c** Illustration of the electronic structure of α-Bi4I4 on the (100) side surface. The gray, yellow, and red curves indicate the bulk, surface, and hinge states, respectively. **d-f** The same as **a-c** except for β-Bi4I4. The green dots in **e** and green curves in **f** represent the topological surface states on the (100) surface of β-Bi4I4. **g** Cross-section image of the (010) surface with atomic resolution measured with scanning transmission electron microscope (STEM). **h** X-ray diffraction measurements of Bi4I4.

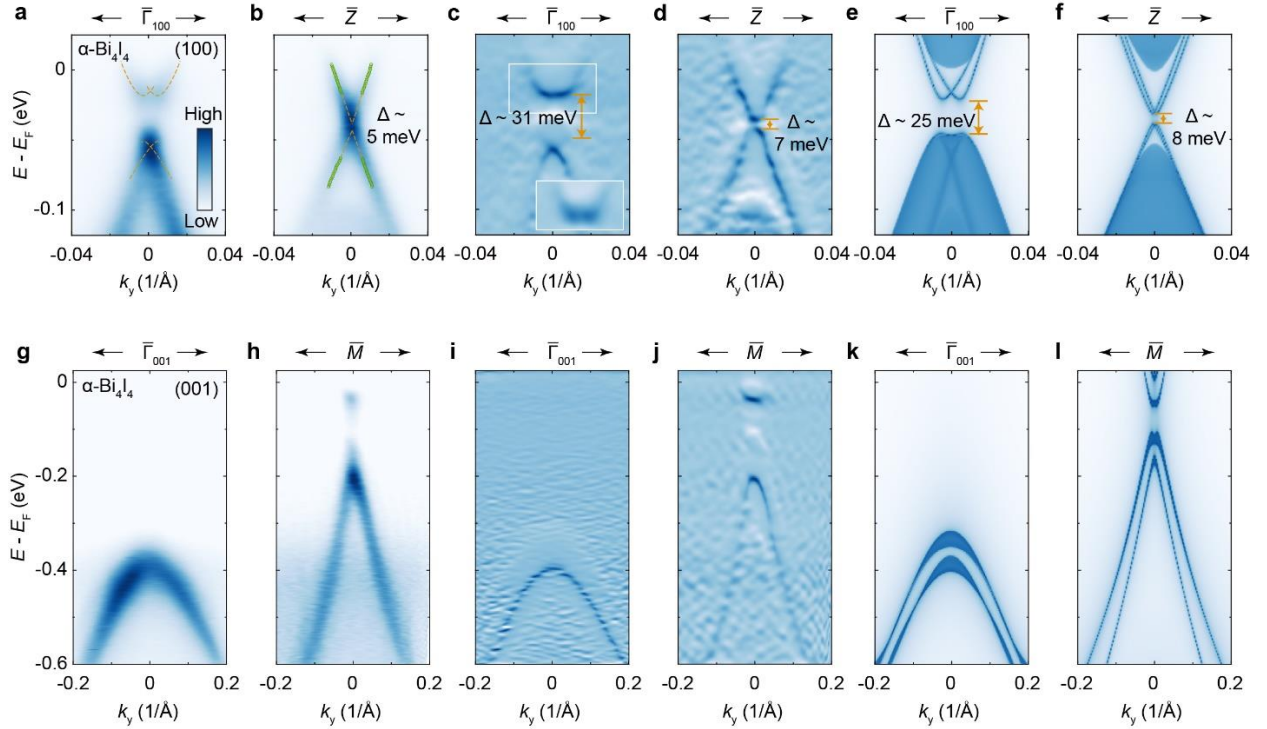

**Fig. 2 Fine electronic structure of α-Bi$_4$I$_4$ on the (001) and (100) surfaces. a**, **b** band structure of the (100) surface states at the $\bar{\Gamma}_{100}$ and $\bar{Z}$ points measured by laser-based micro-ARPES. The yellow dashed lines in **a** are guides to eyes showing the (100) surface states. The green dots in **b** are determined by fitting the momentum distribution curves (MDCs) to Lorentzians and the yellow dashed lines are the fit of the extracted dispersion to the massive Dirac-fermion dispersion. **c**, **d** Curvature plots of the ARPES spectra in **a** and **b**. **e**, **f** (100) surface-projected band structures calculated at the $\bar{\Gamma}_{100}$ and $\bar{Z}$ points, respectively. **g**, **h** ARPES spectra of α-Bi$_4$I$_4$ at the $\bar{\Gamma}_{001}$ and $\bar{M}$ points on the (001) surface. **i**, **j** Curvature plots of the ARPES spectra in **g** and **h**. **k**, **l** (001) surface-projected band structures calculated at the $\bar{\Gamma}_{001}$ and $\bar{M}$ point, respectively. Data were measured along the chain direction at 80 K using LH-polarized 7-eV laser.

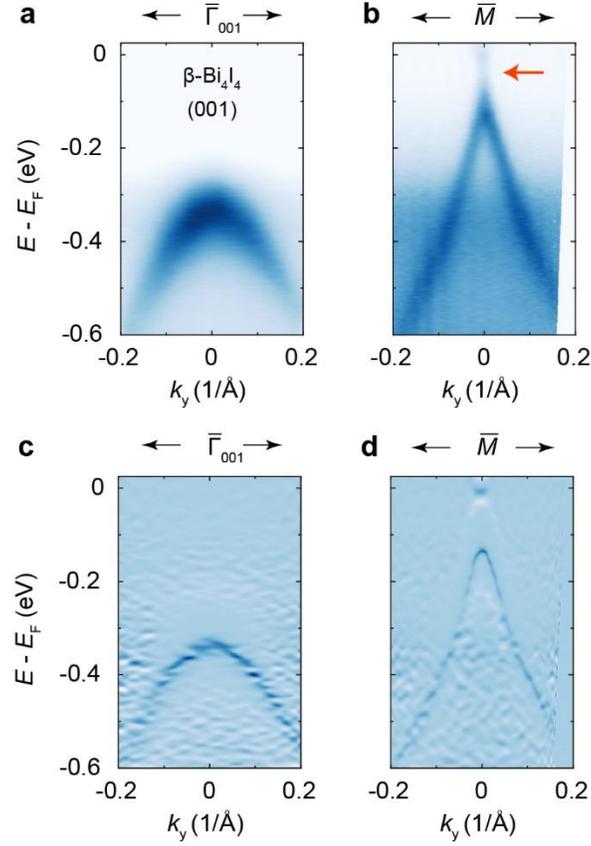

**Fig. 3 Electronic structure of β-Bi$_4$I$_4$ on the (001) surface. a**, **b** ARPES spectra of β-Bi$_4$I$_4$ at $\bar{\Gamma}_{001}$ and $\bar{M}$ point on the (001) surface, respectively. The red arrow indicates the energy gap at the $\bar{M}$ point. **c**, **d** Curvature plot of the ARPES spectra in **a** and **b**. The spectra were measured along the chain direction at 316 K using LH-polarized 7-eV laser.

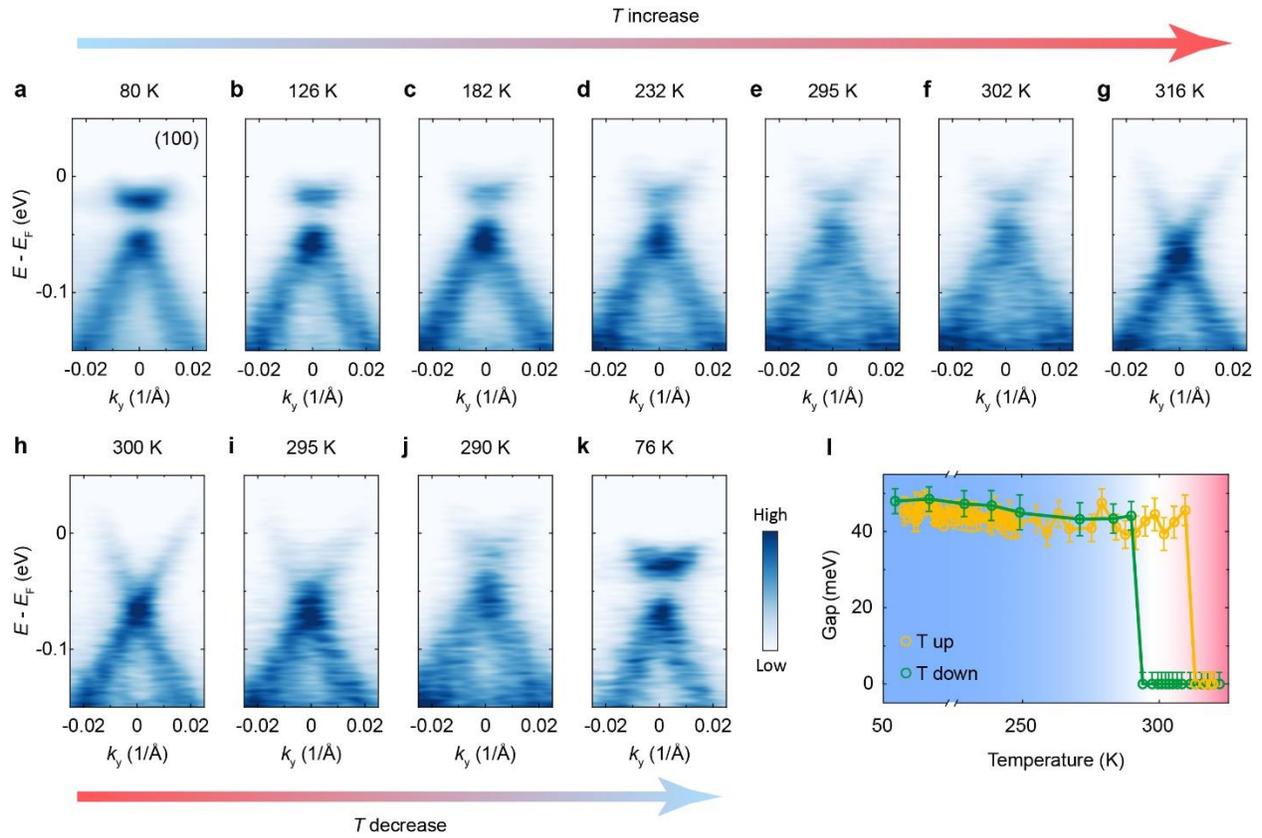

**Fig. 4 Temperature evolution of the (100) surface states across the structural transition. a-k** ARPES spectra of the (100) surface states at the $\bar{\Gamma}_{100}$ point. The measured temperatures are shown above and the arrows indicate the heating or cooling process. **l** Temperature evolution of the (100) surface gap showing a hysteresis loop.